\documentclass[aps,prl,reprint,groupedaddress]{revtex4-1}
\usepackage{graphicx} % Include figure files
\usepackage{dcolumn} % Align table columns on decimal point
\usepackage{bm}% bold math
\usepackage{hyperref}% add hypertext capabilities

\begin{document}

\title{Nuclear dynamics induced by antiprotons}

\author{Zhao-Qing Feng$^{1,2}$}
\email{fengzhq@impcas.ac.cn}
\affiliation{$^{1}$Institute of Modern Physics, Chinese Academy of Sciences, Lanzhou 730000, People's Republic of China  \\
$^{2}$State Key Laboratory of Theoretical Physics and Kavli Institute for Theoretical Physics China, Chinese Academy of Sciences, Beijing 100190, People's Republic of China}

\date{\today}

\begin{abstract}
Reaction dynamics in collisions of antiprotons on nuclei is investigated within the Lanzhou quantum molecular dynamics model. The reaction channels of elastic scattering, annihilation, charge exchange and inelastic collisions of antiprotons on nucleons have been included in the model. Dynamics on particle production, in particular pions, kaons, antikaons and hyperons, is investigated in collisions of $\overline{p}$ on $^{12}$C, $^{20}$Ne, $^{40}$Ca and $^{181}$Ta from a low to high incident momenta. It is found that the annihilations of $\overline{p}$ on nucleons are of importance on the dynamics of particle production in phase space. Hyperons are mainly produced via meson induced reactions on nucleons and strangeness exchange collisions, which lead to the delayed emission in antiproton-nucleus collisions.

\begin{description}
\item[PACS number(s)]
25.43.+t, 24.10.-i, 24.10.Lx
 \item[keywords]
LQMD transport model, antiproton-nucleus collisions, strangeness production
\end{description}
\end{abstract}

\maketitle

\section{I. Introduction}

The dynamics of antiproton-nucleus collisions is a complex process, which is associated with the mean-field potentials of antinucleons and produced particles in nuclear medium, and also with a number of reaction channels, i.e., the annihilation channels, charge-exchange reaction, elastic and inelastic collisions. A more localized energy deposition is able to be produced in antiproton-nucleus collisions in comparison with heavy-ion collisions due to the annihilations. Searching for the cold quark-gluon plasma (QGP) with antiproton beams has been performed as a hot topic both in experiments and in theory calculations over the past several decades. The large yields of strange particles may be produced in antiproton induced reactions, which have the advantage in comparison to proton-nucleus and heavy-ion collisions. The particle production in collisions of antiproton on nuclei has been investigated by using the intranuclear cascade (INC) model \cite{Cl82,Cu89} and the kinetic approach \cite{Ko87}. A number of experimental data were nicely explained within these approaches. Self-consistent description of dynamical evolutions and collisions of antiproton on nucleus with transport models is still very necessary, in particular the fragmentation in collisions of antiproton on nucleus to form hypernuclei.

Strangeness production in antiproton induced nuclear reactions has been investigated thoroughly with the Giessen Boltzmann-Uehling-Uhlenbeck (GiBUU) transport model \cite{Ga12,La12} and the Lanzhou quantum molecular dynamics (LQMD) approach \cite{Fe14}. The production of hypernuclei is associated with the reaction channels of hyperons and also hyperon-nucleon (HN) potential. From comparison of kinetic energy or momentum spectra of hyperons to experimental data, the HN potential can be extracted. Also the antinucleon-nucleon potential is able to be constrained from particle production. The dynamical mechanism on strange particle production can be explored from the analysis of reaction channels and comparison to experimental spectra.

\section{II. Model description}

In the Lanzhou quantum molecular dynamics (LQMD) model, the dynamics of resonances ($\Delta$(1232), N*(1440), N*(1535) etc), hyperons ($\Lambda$, $\Sigma$, $\Xi$) and mesons ($\pi$, $K$, $\eta$, $\overline{K}$, $\rho$, $\omega$ ) is described via hadron-hadron collisions, decays of resonances and mean-field potentials in nuclear medium \cite{Fe09,Fe11}. The evolutions of baryons (nucleons, resonances and hyperons), anti-baryons and mesons in the collisions are governed by Hamilton's equations of motion which read as
\begin{eqnarray}
\dot{\mathbf{p}}_{i}=-\frac{\partial H}{\partial\mathbf{r}_{i}},
\quad \dot{\mathbf{r}}_{i}=\frac{\partial H}{\partial\mathbf{p}_{i}}.
\end{eqnarray}
The Hamiltonian of baryons consists of the relativistic energy, the effective interaction potential and the momentum dependent part as follows:
\begin{equation}
H_{B}=\sum_{i}\sqrt{\textbf{p}_{i}^{2}+m_{i}^{2}}+U_{int}+U_{mom}.
\end{equation}
Here the $\textbf{p}_{i}$ and $m_{i}$ represent the momentum and the mass of the baryons. The effective interaction potential $U_{int}$ is composed of the Coulomb interaction and the local potential \cite{Fe11}. A Skyrme-type momentum dependent interaction has been used in the evaluation of the local potential energy for nucleons and resonances. The effect of the momentum dependence of the symmetry potential in heavy-ion collisions was also investigated with the isospin-dependent Boltzmann Uehling Uhlenbeck transport model \cite{Ga11}.

The hyperon mean-field potential is constructed on the basis of the light-quark counting rule. The self-energies of $\Lambda$ and $\Sigma$ are assumed to be two thirds of that experienced by nucleons. And the $\Xi$ self-energy is one third of nucleon's ones. Thus, the in-medium dispersion relation reads
\begin{equation}
\omega_{B}(\textbf{p}_{i},\rho_{i})=\sqrt{(m_{B}+\Sigma_{S}^{B})^{2}+\textbf{p}_{i}^{2}} + \Sigma_{V}^{B},
\end{equation}
e.g., for hyperons $\Sigma_{S}^{\Lambda,\Sigma}= 2 \Sigma_{S}^{N}/3$, $\Sigma_{V}^{\Lambda,\Sigma}= 2 \Sigma_{V}^{N}/3$, $\Sigma_{S}^{\Xi}= \Sigma_{S}^{N}/3$ and $\Sigma_{V}^{\Xi}= \Sigma_{V}^{N}/3$.
The antibaryon energy is computed from the G-parity transformation of baryon potential as
\begin{equation}
\omega_{\overline{B}}(\textbf{p}_{i},\rho_{i})=\sqrt{(m_{\overline{B}}+\Sigma_{S}^{\overline{B}})^{2}+\textbf{p}_{i}^{2}} + \Sigma_{V}^{\overline{B}}
\end{equation}
with $\Sigma_{S}^{\overline{B}}=\Sigma_{S}^{B}$ and $\Sigma_{V}^{\overline{B}}=-\Sigma_{V}^{B}$.
The nuclear scalar $\Sigma_{S}^{N}$ and vector $\Sigma_{V}^{N}$ self-energies are computed from the well-known relativistic mean-field model with the NL3 parameter ($g_{\sigma N}^{2}$=80.82, $g_{\omega N}^{2}$=155). The optical potential of baryon or antibaryon is derived from the in-medium energy as $V_{opt}(\textbf{p},\rho)=\omega(\textbf{p},\rho)-\sqrt{\textbf{p}^{2}+m_{B}^{2}}$. A factor $\xi$ is introduced in the evaluation of antinucleon optical potential to mimic the antiproton-nucleus scattering \cite{La09} and the real part of phenomenological antinucleon-nucleon optical potential \cite{Co82} as $\Sigma_{S}^{\overline{N}}=\xi\Sigma_{S}^{N}$ and $\Sigma_{V}^{\overline{N}}=-\xi\Sigma_{V}^{N}$ with $\xi$=0.25, which leads to the optical potential $V_{\overline{N}}$=-164 MeV for an antinucleon at the zero momentum and normal nuclear density $\rho_{0}$=0.16 fm$^{-3}$. Shown in Fig. 1 is a comparison of the baryon and antibaryon energies in nuclear medium and the optical potentials as a function of baryon density. The empirical $\Lambda$ potential extracted from hypernuclei experiments \cite{Mi88} is well reproduced with the approach. The antihyperons exhibit strongly attractive potentials in nuclear medium. The optical potentials of hyperons will affect the production of hypernuclei. The values of optical potentials at saturation density are -32 MeV, -16 MeV, -164 MeV, -436 MeV and -218 MeV for $\Lambda(\Sigma)$, $\Xi$, $\overline{N}$, $\overline{\Lambda}$ and $\overline{\Xi}$, respectively.

%%%%%%%%%%%%%%%%%%%%%%%%%%%%%%%%%%%%%%% figure 1 %%%%%%%%%%%%%%%%%%%%%%%%
\begin{figure*}
\includegraphics{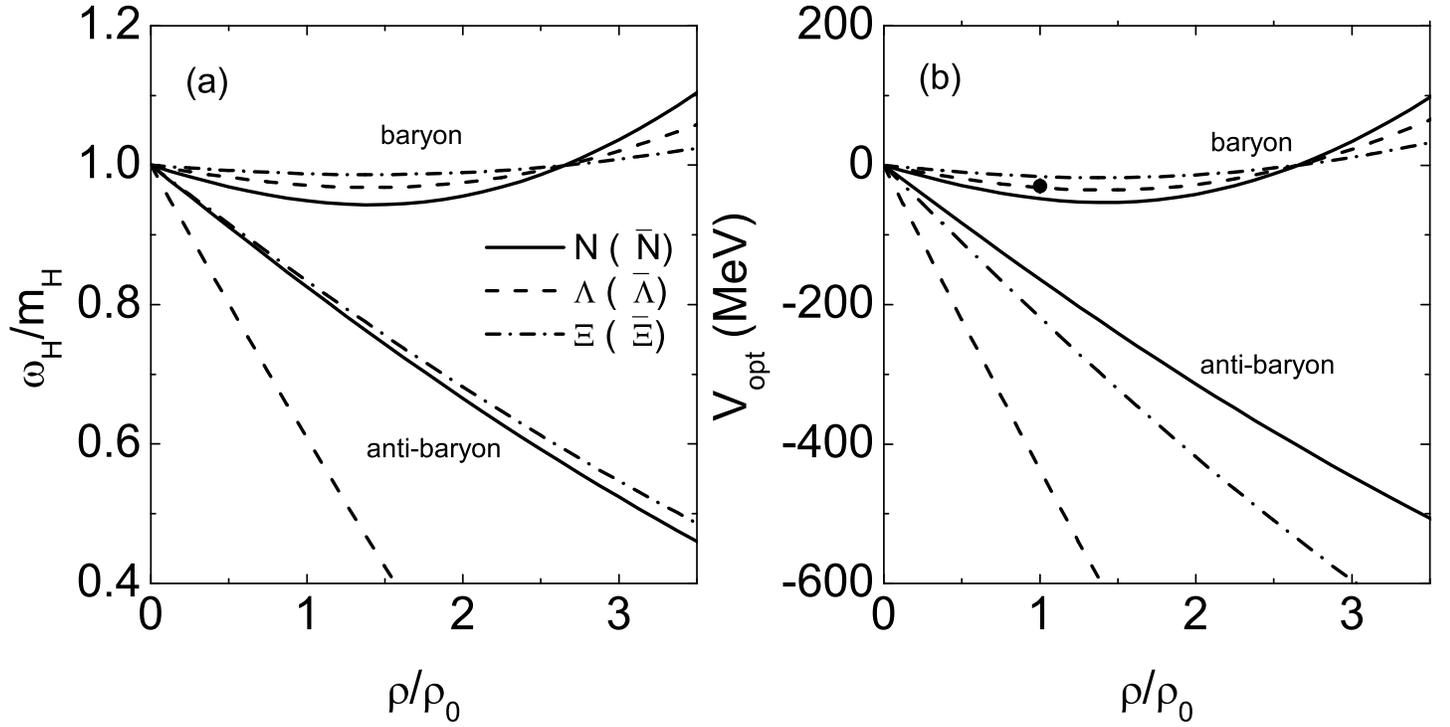}
\caption{\label{fig:wide} Density dependence of the in-medium energies of nucleon, hyperon and their antiparticles in units of free mass and the optical potentials at the momentum of $\textbf{p}$=0 GeV/c. The empirical value of $\Lambda$ potential extracted from hypernuclei experiments is denoted by the solid circle \cite{Mi88}.}
\end{figure*}
%%%%%%%%%%%%%%%%%%%%%%%%%%%%%%%%%%%%%%%%%%%%%%%%%%%%%%%%%%%%%%%%%%%%%%%%%

The evolution of mesons is also determined by the Hamiltonian, which is given by
\begin{eqnarray}
H_{M}&& = \sum_{i=1}^{N_{M}}\left( V_{i}^{\textrm{Coul}} + \omega(\textbf{p}_{i},\rho_{i}) \right).
\end{eqnarray}
Here the Coulomb interaction is given by
\begin{equation}
V_{i}^{\textrm{Coul}}=\sum_{j=1}^{N_{B}}\frac{e_{i}e_{j}}{r_{ij}},
\end{equation}
where the $N_{M}$ and $N_{B}$ are the total numbers of mesons and baryons including charged resonances.The kaon and anti-kaon energies in the nuclear medium distinguish isospin effects based on the chiral Lagrangian approach as \cite{Fe13,Ka86,Sc97}
\begin{eqnarray}
\omega_{K}(\textbf{p}_{i},\rho_{i})= && \left[m_{K}^{2}+\textbf{p}_{i}^{2}-a_{K}\rho_{i}^{S}
-\tau_{3}c_{K}\rho_{i3}^{S}+(b_{K}\rho_{i}+\tau_{3}d_{K}\rho_{i3})^{2}\right]^{1/2}
\nonumber \\
&& +b_{K}\rho_{i}+\tau_{3}d_{K}\rho_{i3}
\end{eqnarray}
and
\begin{eqnarray}
\omega_{\overline{K}}(\textbf{p}_{i},\rho_{i})= && \left[m_{\overline{K}}^{2}+\textbf{p}_{i}^{2}-a_{\overline{K}}\rho_{i}^{S}
-\tau_{3}c_{K}\rho_{i3}^{S}+(b_{K}\rho_{i}+\tau_{3}d_{K}\rho_{i3})^{2}\right]^{1/2}
\nonumber \\
&& -b_{K}\rho_{i}-\tau_{3}d_{K}\rho_{i3},
\end{eqnarray}
respectively. Here the $b_{K}=3/(8f_{\pi}^{\ast 2})\approx$0.333 GeVfm$^{3}$, the $a_{K}$ and $a_{\overline{K}}$ are 0.18 GeV$^{2}$fm$^{3}$ and 0.31 GeV$^{2}$fm$^{3}$, respectively, which result in the strengths of repulsive kaon-nucleon (KN) potential and of attractive antikaon-nucleon $\overline{K}$N potential with the values of 27.8 MeV and -100.3 MeV at saturation baryon density for isospin symmetric matter, respectively. The $\tau_{3}$=1 and -1 for the isospin pair K$^{+}$($\overline{K}^{0}$) and K$^{0}$(K$^{-}$), respectively. The parameters $c_{K}$=0.0298 GeV$^{2}$fm$^{3}$ and $d_{K}$=0.111 GeVfm$^{3}$ determine the isospin splitting of kaons in neutron-rich nuclear matter. The optical potential of kaon is derived from the in-medium energy as $V_{opt}(\textbf{p},\rho)=\omega(\textbf{p},\rho)-\sqrt{\textbf{p}^{2}+m_{K}^{2}}$. The values of $m^{\ast}_{K}/m_{K}$=1.056 and $m^{\ast}_{\overline{K}}/m_{\overline{K}}$=0.797 at normal baryon density are concluded with the parameters in isospin symmetric nuclear matter. The effective mass is used to evaluate the threshold energy for kaon and antikaon production, e.g., the threshold energy in the pion-baryon collisions $\sqrt{s_{th}}=m^{\ast}_{Y} + m^{\ast}_{K}$.

Based on hadron-hadron collisions in heavy-ion reactions in the region of 1-2 A GeV energies \cite{Fe11}, we have further included the annihilation channels, charge-exchange reaction, elastic and inelastic scattering in antinucleon-nucleon collisions: $\overline{N}N \rightarrow \overline{N}N$, $\overline{N}N \rightarrow \overline{B}B$,
$\overline{N}N \rightarrow \overline{Y}Y$ and $\overline{N}N \rightarrow \texttt{annihilation}(\pi,\eta,\rho,\omega,K,\overline{K},K^{\ast},\overline{K}^{\ast},\phi)$.
Here the B strands for (N, $\triangle$, N$^{\ast}$) and Y($\Lambda$, $\Sigma$, $\Xi$), K(K$^{0}$, K$^{+}$) and $\overline{K}$($\overline{K^{0}}$, K$^{-}$). The overline of B (Y) means its antiparticle. The cross sections of these channels are based on the parametrization of experimental data \cite{Bu12}. The $\overline{N}N$ annihilation is described by a statistical model with SU(3) symmetry \cite{Go92}, which includes various combinations of possible emitted mesons with the final state up to six particles \cite{La12}. A hard core scattering is assumed in two-particle collisions by Monte Carlo procedures, in which the scattering of two particles is determined by a geometrical minimum distance criterion $d\leq\sqrt{\sigma_{tot}/\pi}$ fm weighted by the Pauli blocking of the final states. Here, the total cross section $\sigma_{tot}$ in mb is the sum of all possible channels. The probability reaching a channel in a collision is calculated by its contribution of the channel cross section to the total cross section as $P_{ch}=\sigma_{ch}/\sigma_{tot}$. The choice of the channel is done randomly by the weight of the probability.

\section{III. Results and discussions}

The emission mechanism of particles produced in antiproton induced reactions is significant in understanding contributions of different reaction channels associated with antiprotons on nucleons and secondary collisions. Shown in Fig. 2 the temporal evolutions of pions, kaons, anti-kaons and hyperons in the reaction of $\overline{p}$+$^{12}$C at an incident momentum of 1 GeV/c. It is shown that the kaons are emitted immediately after the annihilation of antinucleons and nucleons. The secondary collisions of pions and anti-kaons on nucleons retard the emissions in the reaction dynamics, i.e., $\pi N\rightarrow \Delta$, $\overline{K}N\rightarrow \pi Y$ etc, which lead to the production of hyperons. At the considered momentum below its threshold energy, e.g., the reaction $\overline{N}N\rightarrow \overline{\Lambda}\Lambda$ (p$_{threshold}$=1.439 GeV/c), hyperons are mainly contributed from the secondary collisions after annihilations, i.e., $\pi N\rightarrow KY$ and $\overline{K}N\rightarrow \pi Y$.

%%%%%%%%%%%%%%%%%%%%%%%%%%%%%%%%%%%%%%% figure 2 %%%%%%%%%%%%%%%%%%%%%%%%
\begin{figure*}
\includegraphics{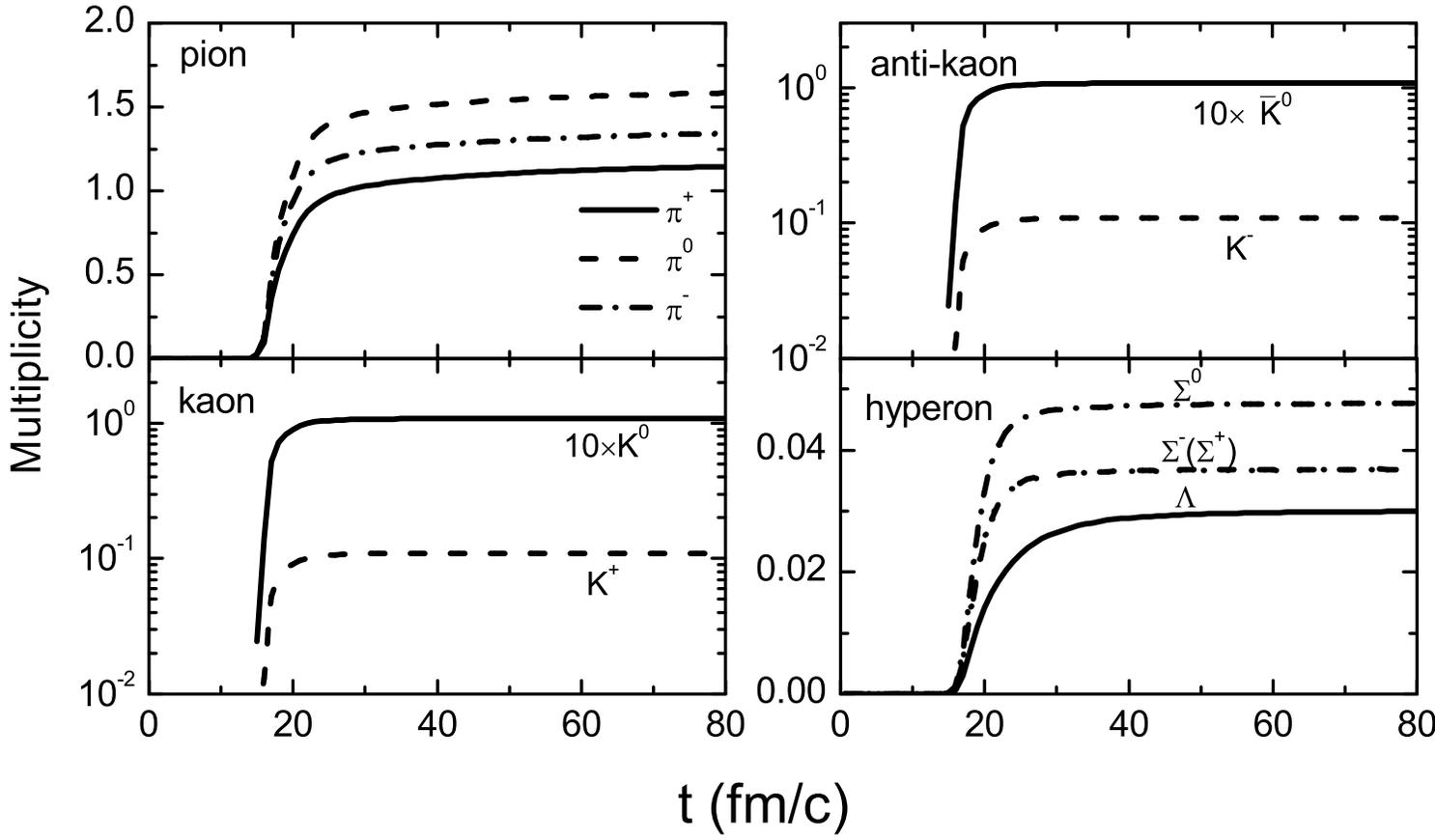}
\caption{\label{fig:wide} Time evolutions of particles produced in collisions of $\overline{p}$ on $^{12}$C at incident momentum of 1 GeV/c.}
\end{figure*}
%%%%%%%%%%%%%%%%%%%%%%%%%%%%%%%%%%%%%%%%%%%%%%%%%%%%%%%%%%%%%%%%%%%%%%%%%

Phase-space distributions of particles produced in heavy-ion collisions were used to investigate the in-medium properties of hadrons, in particular for strange particles \cite{Fe13}. It has been shown that the mean-field potentials influence the spectrum structures, i.e., the kinetic energy spectra of inclusive cross sections, rapidity and transverse momentum distributions of particles etc. The phase-space structure of particle emission in antiproton induced reactions is also investigated in this work. Shown in Fig. 3 is the transverse momentum distributions of neutral particles in collisions of antiprotons on different targets. The hyperon emission is coupled to the strangeness exchange reactions. The yields increase with the mass number of target. Shown in Fig. 4 is the rapidity distributions. Target effect is pronounced in the domain of antiproton-nucleon frame (y=0.46) for $\pi^{0}$, $K^{0}$ and $\overline{K}^{0}$ emission, where heavier target increases the production of pions and kaons. It leads to an opposite contribution for antikaon production because of strangeness exchange $\overline{K}N\rightarrow \pi Y$. The peak of hyperon emission is close to the target rapidity and more sensitive to the mass .

%%%%%%%%%%%%%%%%%%%%%%%%%%%%%%%%%%%%%%% figure 3 %%%%%%%%%%%%%%%%%%%%%%%%
\begin{figure*}
\includegraphics{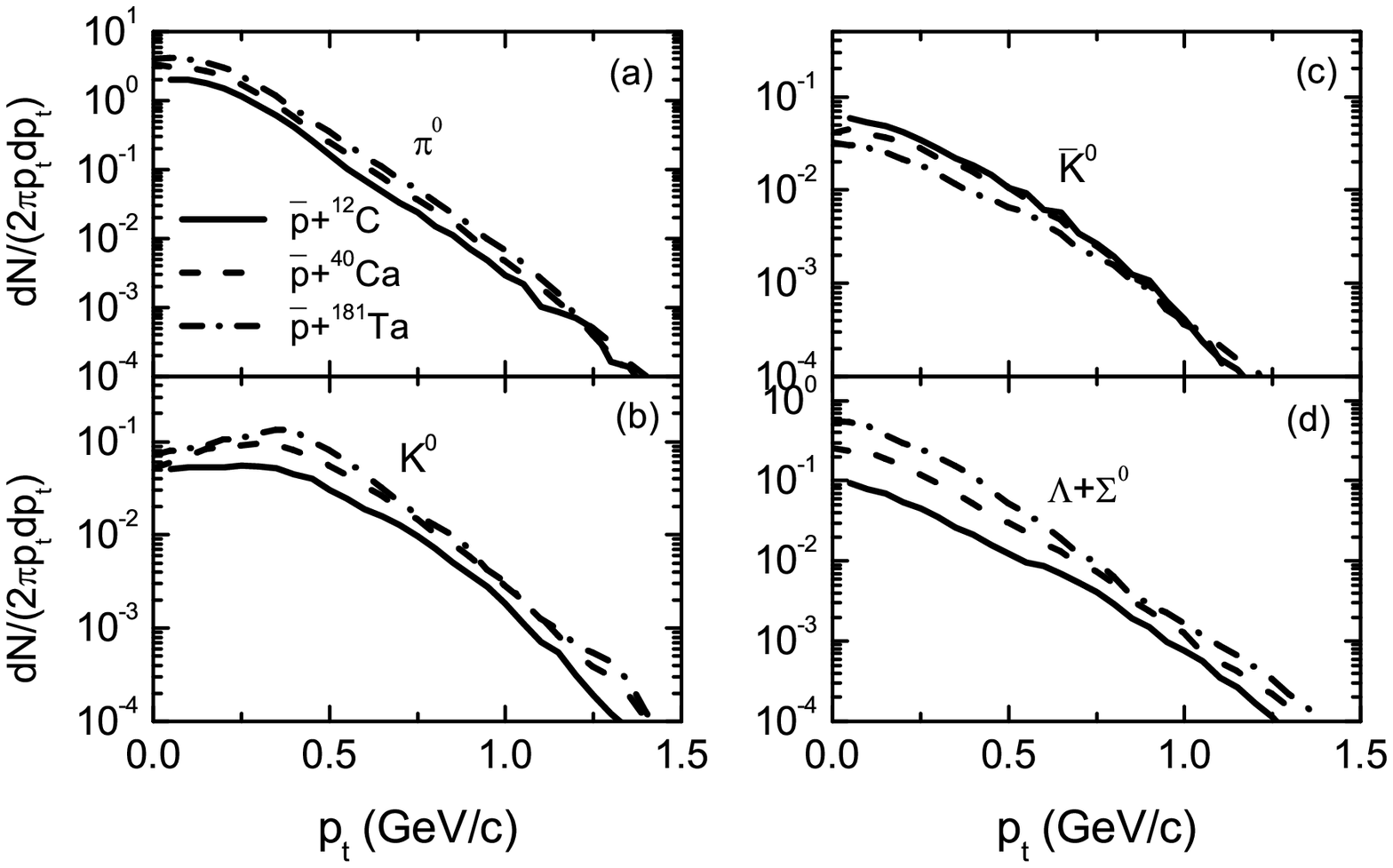}
\caption{\label{fig:wide} The transverse momentum distributions of $\pi$, K, $\overline{K}$ and neutral hyperons produced in $\overline{p}$ induced reactions on $^{12}$C, $^{20}$Ne, $^{40}$Ca and $^{181}$Ta at incident momentum of 4 GeV/c.}
\end{figure*}
%%%%%%%%%%%%%%%%%%%%%%%%%%%%%%%%%%%%%%%%%%%%%%%%%%%%%%%%%%%%%%%%%%%%%%%%%

%%%%%%%%%%%%%%%%%%%%%%%%%%%%%%%%%%%%%%% figure 4 %%%%%%%%%%%%%%%%%%%%%%%%
\begin{figure*}
\includegraphics{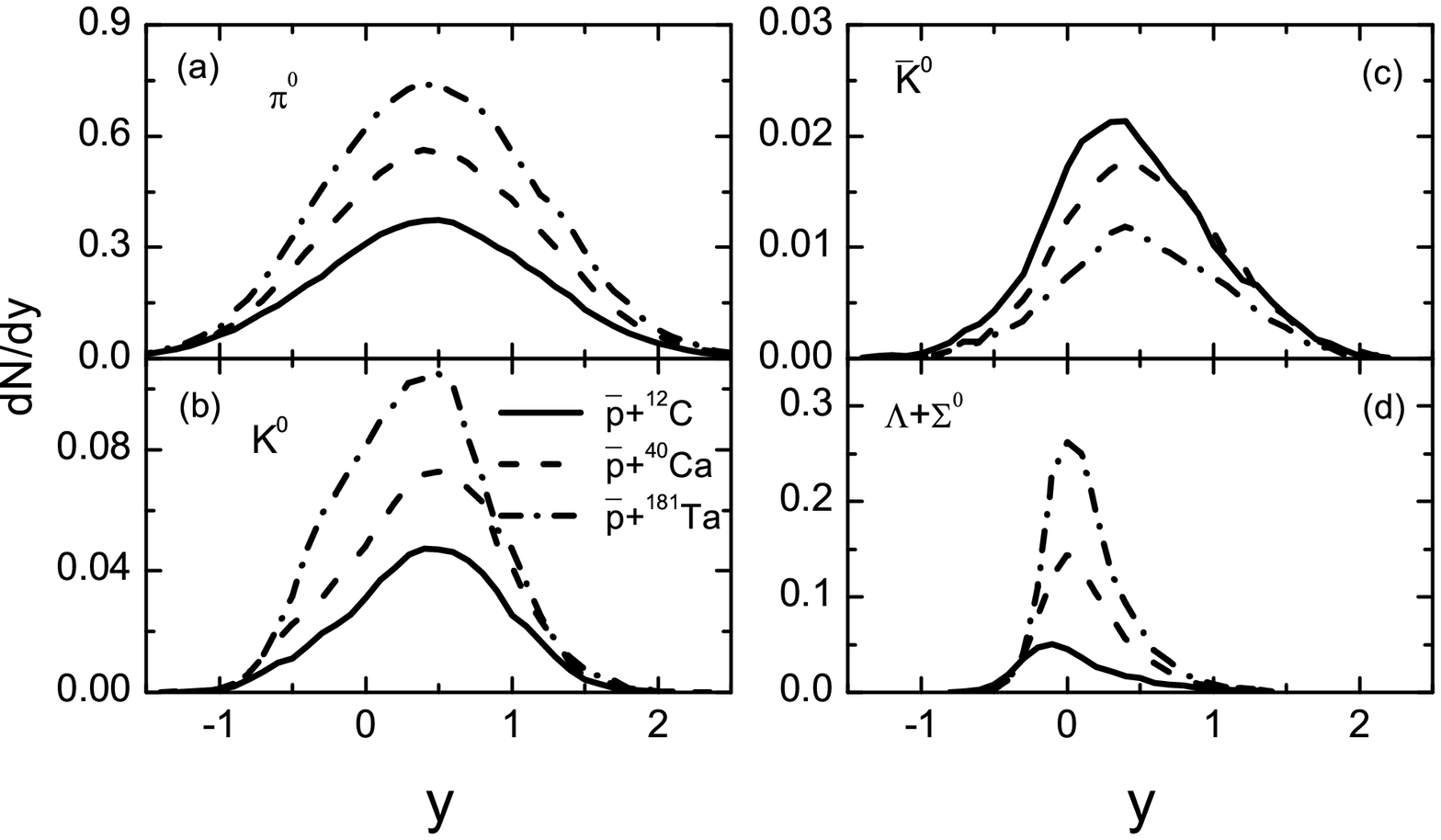}
\caption{\label{fig:wide} The same as in Fig. 3, but for rapidity distributions.}
\end{figure*}
%%%%%%%%%%%%%%%%%%%%%%%%%%%%%%%%%%%%%%%%%%%%%%%%%%%%%%%%%%%%%%%%%%%%%%%%%

\section{IV. Conclusions}

Dynamics on particle production in antiproton induced reactions, in particular pions, kaons, antikaons and hyperons, has been investigated within the LQMD model. The production of pions, kaons and antikaons are mainly contributed from the annihilations of antiproton on nucleons. Hyperons are dominated via the meson-nucleon collisions and strangeness exchange reactions. Kaons are emitted promptly after production in nuclear dynamics. Secondary collisions retard the emission of pions, antikaons and hyperons, which reduce the yields of antikaons and enhance hyperons with increasing the mass number of target.

\textbf{Acknowledgements}

This work was supported by the Major State Basic Research Development Program in China (Nos 2014CB845405 and 2015CB856903), the National Natural Science Foundation of China Projects (Nos 11175218 and U1332207) and the Youth Innovation Promotion Association of Chinese Academy of Sciences.

\end{document}